\def\ra{\rangle}
\def\la{\langle}
\def\be{\begin{equation}}
\def\ee{\end{equation}}
\def\ba{\begin{array}}
\def\ea{\end{array}}

\documentclass[prl,showpacs,preprint,amsmath,amssymb]{revtex4}
\usepackage{graphicx}
\usepackage{caption2}
\usepackage{mathrsfs,graphicx,color,float,indentfirst,textcomp}
\usepackage{latexsym,lscape,enumerate}
\def\qed{\leavevmode\unskip\penalty9999 \hbox{}\nobreak\hfill
     \quad\hbox{\leavevmode  \hbox to.77778em{%
               \hfil\vrule   \vbox to.675em%
               {\hrule width.6em\vfil\hrule}\vrule\hfil}}
     \par\vskip3pt}

\begin{document}
\title{Trade-off Relations of Bell Violations among Pairwise Qubit Systems}
\author{Hui-Hui Qin$^{1}$}
\author{Shao-Ming Fei$^{2,3}$}
\author{Xianqing Li-Jost$^{3}$}

\affiliation{$^1$Department of Mathematics, School of Science, South
China University of Technology, Guangzhou 510640, China\\
$^2$School of Mathematical Sciences, Capital Normal University,
Beijing 100048, China\\
$^3$Max-Planck-Institute for Mathematics in the Sciences, 04103
Leipzig, Germany}

\begin{abstract}

We investigate the non-locality distributions among multi-qubit systems based on the
maximal violations of the CHSH inequality of the reduced pairwise qubit systems.
We present a trade-off relation satisfied by these maximal violations, which
gives rise to restrictions on the distribution of non-locality among the
sub-qubit systems. For a three-qubit system, it is impossible that all pair of qubits violate the CHSH inequality,
and once a pair of qubits violates the CHSH inequality maximally, the other two pairs of qubits
must both obey the CHSH inequality. Detailed examples are given to display the trade-off relations, and
the trade-off relations are generalized to arbitrary multi-qubit systems.
\end{abstract}


\maketitle

Quantum mechanics exhibits the nonlocality of the nature, as revealed by the
violation of Bell inequality \cite{Johns}.
A quantum state is said to admit a local
hidden variable (LHV) model if all the measurement outcomes
can be modeled as a classical random distribution over
a probability space. All quantum states
admitting LHV models satisfy any Bell inequalities.
A state that admits no LHV models must violate at least one Bell inequality.
The Bell inequality provides the way to distinguish experimentally between quantum mechanical
predictions and predictions of local realistic models.
The violation of the Bell inequalities is also closely related
to the extraordinary power of realizing certain tasks in quantum
information processing such as quantum protocols to decrease communication
complexity \cite{commun} and secure quantum communication \cite{secure}.

For pure quantum states, the quantum entanglement coincides with the violation
of Bell inequalities. Namely, for pure states the entanglement and the non-locality coincide.
Any pure entangled states violate
a Bell inequality \cite{gisin,gisinperes,PopescuRohrlich92,jingling,limingbell,yusixia}.
However, a general mixed entangled state could admit LHV models.
There has been no effective method to judge whether
a mixed state admits a LHV model or not \cite{Pitowsky,AlonNaor,liming}.
Even for the simple two-qubit Werner states,
the precise threshold value of nonlocality is still unknown \cite{huabobo}.

As one of the fundamental differences between quantum entanglement and classical
correlations, a key property of entanglement is that a quantum system entangled with one of other
systems limits its entanglement with the remaining ones.
The monogamy relations give rise to the distribution of quantum entanglement in
a multipartite systems \cite{022309,PRA80044301,C2,E1,zhuxn}. Monogamy is also an essential feature
allowing for security in quantum key distribution \cite{k3}.

An interesting question one may ask is what the distribution of non-locality
in a multipartite system would be. Namely, would a quantum system that has non-local correlations with one of other
systems limit its non-local correlations with the remaining systems?
In this paper, by using the Clauser-Horne-Shimony-Holt (CHSH) inequality \cite{clauser},
we study such non-locality distributions among the multi-qubit systems.
We show that quantum correlations captured by the violation of Bell inequalities have to obey interesting trade-off
relations, similar to the distribution of quantum entanglement in multipartite systems.
We present the analytical trade-off relations obeyed by the CHSH test of pairwise qubits
in a three-quibt system. The result is then generalized to general multi-qubit systems.

The well-known CHSH \cite{clauser} inequality is feasible for experimental verifications.
Suppose two observers, Alice and Bob, are separated spatially and
share two qubits. Alice and Bob each measures a dichotomic observable
with possible outcomes $\pm 1$ in one of two measurement settings:
$A_1, A_2$ and $B_1, B_2$, respectively. The CHSH inequality is a
constraint on the correlations between Alice's and Bob's measurement
outcomes if a local realistic description is assumed. The
the corresponding Bell operator is given by
\be\label{chsh22}
\mathcal{B}=A_1\otimes B_1+A_1\otimes
B_2+A_2\otimes B_1-A_2\otimes B_2,
\ee
where $A_i=\vec{a}_i\cdot\vec{\sigma}_A=a_i^x\sigma_A^1+a_i^y\sigma_A^2+a_i^z\sigma_A^3$,
$B_j=\vec{b}_j\cdot \vec{\sigma}_B=b_j^x\sigma_B^1+b_j^y\sigma_B^2+b_j^z\sigma_B^3$,
$\vec{a}_i=(a_i^x,a_i^y,a_i^z)$ and $\vec{b}_j=(b_j^x,b_j^y,b_j^z)$
are real unit vectors satisfying $|\vec{a}_i|=|\vec{b}_j|=1$ with
$i,j=1,2$, and $\sigma_{A/B}^{1,2,3}$ are Pauli matrices. The CHSH
inequality says that if there exist LHV models to describe the system, the inequality
$|\la{\mathcal {B}} \ra_{\rho}|\leq2$
must hold, where $\la{\mathcal {B}} \ra=tr(\rho\mathcal{B})$ is the mean value
of the Bell operator $\mathcal{B}$ associated with the system state $\rho$.
For quantum entangled pure states, it is always possible to find suitable
observables $A_1$, $A_2$, $B_1$ and $B_2$ such that inequality $|\la{\mathcal {B}} \ra_{\rho}|\leq2$
is violated. For instance, taking the maximally entangled state
$|\psi_+\ra=(|01\ra-|10\ra)/\sqrt{2}$, one may set $A_1=\sigma^1$,
$A_2=\sigma^3$, $B_1=(\sigma^1+\sigma^3)/\sqrt{2}$, and
$B_2=(\sigma^1-\sigma^3)/\sqrt{2}$. Then one gets $|\la{\mathcal
{B}}\ra|=2\sqrt{2}$, which gives the maximal violation of the CHSH inequality \cite{Nielsen}.

A two-qubit quantum state $\rho$ can be always expressed in terms of Pauli
matrices $\sigma_i$, $i=1,2,3$,
\be\label{rho}
\rho=\frac{1}{4}
I\otimes I +\sum_{i=1}^3 r_i\,\sigma_i\otimes I+\sum_{j=1}^3
s_j\, I\otimes\sigma_j +\sum_{i,j=1}^3 m_{ij}\,\sigma_i\otimes \sigma_j,
\ee
where $I$ is the $2\times 2$ identity matrix, $r_k=\frac{1}{4}tr(\rho\,\sigma_k\otimes I)$,
$s_l=\frac{1}{4}tr(\rho\, I\otimes\sigma_l)$ and
$m_{kl}=\frac{1}{4}tr(\rho\,\sigma_k\otimes\sigma_l)$. We denote $M$ the matrix with entries $m_{ij}$.
Let $\langle CHSH \rangle_{\rho}$ denote the maximal mean value
$\langle\mathcal{B}\rangle_{\rho}$ under all possible measurement settings $\{A_{i},B_{j}\}$.
Then for a given two-qubit state $\rho$, $\langle CHSH \rangle_{\rho}$ is given by \cite{Horodecki},
\begin{equation} \label{eq-1}
\langle CHSH \rangle_{\rho}=\max_{A_{i},B_{j}}tr(\rho\mathcal{B})=2\sqrt{\tau_{1}+\tau_{2}},
\end{equation}
where $\tau_{1}$, $\tau_{2}$ are the two largest eigenvalues of the matrix $M^{\dag}M$, $M^{\dag}$ is the
conjugate and transpose of the $3\times3$ matrix $M$.

We first consider three-qubit systems. Let $\mathbf{H}^{A}$, $\mathbf{H}^{B}$ and $\mathbf{H}^{C}$
be two-dimensional Hilbert spaces. For a three-qubit state
$\rho_{ABC} \in \mathbf{H}^{A} \otimes \mathbf{H}^{B} \otimes \mathbf{H}^{C}$, we denote
$\rho_{AB}=tr_C \rho_{ABC}$, $\rho_{AC}=tr_B \rho_{ABC}$, $\rho_{BC}=tr_A \rho_{ABC}$ the reduced two-qubit density matrices of $\rho_{ABC}$.

{\bf Theorem}. For any three-qubit state
$\rho_{ABC} \in \mathbf{H}^{A} \otimes \mathbf{H}^{B} \otimes \mathbf{H}^{C}$,
the maximal violation of the CHSH tests on the pairwise bipartite states satisfies the following trade-off relation,
\begin{equation}\label{eq1}
\langle CHSH\rangle^{2}_{\rho_{AB}}+ \langle CHSH \rangle^{2}_{\rho_{AC}} + \langle CHSH\rangle^{2}_{\rho_{BC}} \leq 12.
\end{equation}

{\sf Proof}. From the formula (\ref{eq-1}), for any bipartite state $\rho_{XY}$ $(XY=AB, AC, BC)$, the square of its maximal value of
the CHSH Bell operator satisfies
\begin{equation}\label{pp}
\langle CHSH \rangle^{2}_{\rho_{XY}}=4(\sum_{s,t}m^{XY}_{st}(m^{XY}_{st})^{*}-\min{\tau}) \leq 4\sum_{s,t}m^{XY}_{st}(m^{XY}_{st})^{*},
\end{equation}
where $\min{\tau}$ is the minimum eigenvalue of $M^{\dag}M$, $m^{XY}_{st}$ are the entries of the corresponding matrix $M$
with respect to the state $\rho_{XY}$.

We first prove the Theorem for the case of pure states.
Consider a pure three-qubit state $|\psi\rangle =\sum^{1}_{i,j,k=0} a_{ijk}|ijk \rangle \in \mathbf{H}^{A}\otimes \mathbf{H}^{B}\otimes \mathbf{H}^{C}$,
where $a_{ijk}$ satisfy the normalization condition, $\sum^{1}_{i,j,k=0} a_{ijk}a^{*}_{ijk}=1$.
From (\ref{pp}) we have
$$
\begin{aligned}
\langle CHSH \rangle^{2}_{\rho_{AB}}+\langle CHSH \rangle^{2}_{\rho_{AC}}+\langle CHSH \rangle^{2}_{\rho_{BC}} \\
\leq
4\sum^{3}_{s,t=1}[m^{AB}_{st}(m^{AB}_{st})^{*}+m^{AC}_{st}(m^{AC}_{st})^{*}+m^{BC}_{st}(m^{BC}_{st})^{*}],
\end{aligned}
$$
where $\rho_{AB}=tr_{C}(|\psi\rangle \langle \psi|)=
\sum^{1}_{i_{m},j_{m}=0}\sum^{1}_{i_{n},j_{n}=0}\sum^{1}_{k=0}a_{i_{m}j_{m}k}a^\ast_{i_{n}j_{n}k} |i_{m}j_{m}\rangle \langle i_{n}j_{n}|\in
\mathbf{H}^{A}\otimes \mathbf{H}^{B}$, $m^{AB}_{st}=tr(\rho_{AB}\sigma_{s}\otimes \sigma_{t})$. So do $\rho_{AC}$, $m^{AC}_{st}$ and $\rho_{BC}$, $m^{BC}_{st}$.
From the explicit expressions of the coefficients $m^{AB}_{st}$, $m^{AC}_{st}$ and $m^{BC}_{st}$ we have
$$
\begin{array}{l}
\displaystyle \sum^{3}_{s,t=1} [m^{AB}_{st}(m^{AB}_{st})^{*}+m^{AC}_{st}(m^{AC}_{st})^{*}+m^{BC}_{st}(m^{BC}_{st})^{*}]\\[5mm]
\displaystyle = \sum_{1\leq k<l \leq3}\{1+\sum_{i_{k} \neq j_{k}}\sum_{i_{l},j_{l}}\sum_{i_{kl},j_{kl}}[a_{i_{k}i_{l}i_{kl}}
a_{j_{k}j_{l}j_{kl}}(a_{i_{k}i_{l}j_{kl}})^{*}(a_{j_{k}j_{l}i_{kl}})^{*}
-a_{i_{k}i_{l}i_{kl}}a_{j_{k}j_{l}j_{kl}}(a_{i_{k}j_{l}i_{kl}})^{*}(a_{j_{k}i_{l}j_{kl}})^{*}]\\[6mm]
\displaystyle \qquad +\sum_{i_{l}\neq j_{l}}\sum_{i_{k},j_{k}}\sum_{i_{kl},j_{kl}}[a_{i_{k}i_{l}i_{kl}}a_{j_{k}j_{l}j_{kl}}
(a_{i_{k}i_{l}j_{kl}})^{*}(a_{j_{k}j_{l}i_{kl}})^{*}
-a_{i_{k}i_{l}i_{kl}}a_{j_{k}j_{l}j_{kl}}(a_{i_{k}j_{l}j_{kl}})^{*}(a_{j_{k}i_{l}i_{kl}})^{*}]\} \\[5mm]
\displaystyle =\sum_{1\leq k <l \leq 3}1=3.
\end{array}
$$
Therefore for any pure three-qubit state, we have the trade-off relation (\ref{eq1}).

Now for any mixed state $\rho_{ABC}=\Sigma_{i}\,p_{i}|\phi_{i}\rangle\langle \phi_{i}|\in \mathbf{H}^{A}\otimes \mathbf{H}^{B}\otimes \mathbf{H}^{C}$,
$0<p_{i}\leq 1$, $\sum_{i}p_{i}=1$, we have
$$
\begin{aligned}
\displaystyle &\frac{1}{4}(\langle CHSH \rangle^{2}_{tr_{A}(\rho_{ABC})}+\langle CHSH \rangle^{2}_{tr_{B}(\rho_{ABC})}+\langle CHSH \rangle^{2}_{tr_{C}(\rho_{ABC})})\\
\displaystyle\quad &\leq(\sum_{i}p_{i}a_{i})^{2}+(\sum_{i}p_{i}b_{i})^{2}+(\sum_{i}p_{i}c_{i})^{2}\\
\displaystyle \quad&\leq\frac{1}{2}\sum_{i,j}p_{i}p_{j}(a^{2}_{i}+b^{2}_{i}+c^{2}_{i}+a^{2}_{j}+b^{2}_{j}+c^{2}_{j})\\
\displaystyle \quad&\leq3\sum_{i,j}p_{i}p_{j}=3,
\end{aligned}
$$
where $a_{i}=\langle CHSH \rangle_{tr_{A}(|\phi_{i}\rangle\langle\phi_{i}|)}$, $b_{i}=\langle CHSH \rangle_{tr_{B}|\phi_{i}\rangle\langle\phi_{i}|)}$,
and $c_{i}=\langle CHSH \rangle_{tr_{C}(|\phi_{i}\rangle\langle\phi_{i}|)}$. Hence (\ref{eq1}) holds also for mixed states. \qed

The trade-off relation (\ref{eq1}) gives rise to restrictions on the distribution of non-locality among the
subsystems. Generally the maximal mean value of the CHSH Bell operator could be $2\sqrt{2}$, i.e.
$\langle CHSH\rangle^{2}_{\rho_{XY}}$ may be 8. However, instead of 24, the bound in the
right hand side of (\ref{eq1}) is 12.

Theorem implies that in a three-qubit system, it is impossible that all the pairs of qubits violate the CHSH inequality,
that is, $\langle CHSH\rangle_{\rho_{XY}}>2$ for all $XY=AB, AC, BC$ would not happen.
Moreover, if one of the three pairs of qubits reaches the maximal violation of the CHSH inequality,
say, $\langle CHSH \rangle_{\rho_{AB}}=2\sqrt{2}$,
then the other two pairs of qubits can not violate the CHSH inequality any more, since in this case we have
$\langle CHSH \rangle^{2}_{\rho_{BC}}+\langle CHSH \rangle^{2}_{\rho_{AC}} \leq 4$, which implies that
$\langle CHSH \rangle_{\rho_{BC}}\leq 2$ and $\langle CHSH \rangle_{\rho_{AC}}\leq 2$.

For an intuitive analysis of the trade-off relation (\ref{eq1}), let us consider the
generalized Schmidt decomposition of a three-qubit state $| \Psi\rangle$ \cite{Acin},
\begin{equation}
|\Psi\rangle=\lambda_{0}|000\rangle+\lambda_{1}e^{i\psi}|100\rangle+\lambda_{2}|101\rangle+\lambda_{3}|110\rangle+\lambda_{4}|111\rangle
\end{equation}
with normalization $\sum_{i}\lambda^{2}_{i}=1$ and $0\leq\psi \leq \pi$.
From the reduced density matix
\begin{equation} \nonumber
\begin{aligned}
\displaystyle&\rho_{AB}=tr_{C}(|\Psi\rangle \langle\Psi|)=
\begin{pmatrix}
\lambda^{2}_{0}&0&\lambda_{0}\lambda_{1}e^{-i\psi}&\lambda_{0}\lambda_{3}\\
0&0&0&0\\
\lambda_{0}\lambda_{1}e^{i\psi}&0&\lambda^{2}_{1}+\lambda^{2}_{2}&\lambda_{1}\lambda_{3}e^{i\psi}+\lambda_{2}\lambda_{4}\\
\lambda_{0}\lambda_{3}&0&\lambda_{1}\lambda_{3}e^{-i\psi}+\lambda_{3}\lambda_{4}&\lambda^{2}_{3}+\lambda^{2}_{4}\\
\end{pmatrix},
\end{aligned}
\end{equation}
one has the corresponding Pauli coefficient matrices $M^{AB}=tr(\rho_{AB}\sigma_{s}\otimes \sigma_{t})$,
$$
\begin{aligned}
M^{AB}=\begin{pmatrix}
2\lambda_{0}\lambda_{3}&0&2\lambda_{0}\lambda_{1}\cos \psi\\
0&-2\lambda_{0}\lambda_{3}&2\lambda_{0}\lambda_{1}\sin \psi\\
-2(\lambda_{1}\lambda_{3}\cos\psi+\lambda_{2}\lambda_{4})&2\lambda_{1}\lambda_{3}\sin\psi&
\lambda^{2}_{0}+\lambda^{2}_{3}+\lambda^{2}_{4}-\lambda^{2}_{1}-\lambda^{2}_{2}
\end{pmatrix}.
\end{aligned}
$$
$M^{AC}$ and $M^{BC}$ can be obtained similarly.

For the simplicity we take $\lambda_{4}=0$. Direct calculation gives
\begin{equation} \label{eq3}
\begin{array}{rcl}
\langle CHSH \rangle^{2}_{\rho_{BC}}&=&2[(1-2\lambda^{2}_{0})^{2}+4(\lambda^{2}_{0}\lambda^{2}_{1}+
 3\lambda^{2}_{2}\lambda^{2}_{3})\\
&&+\sqrt{[(1-2\lambda^{2}_{0})^{2}+4(\lambda^{2}_{0}\lambda^{2}_{1}+
 \lambda^{2}_{2}\lambda^{2}_{3})]^{2}-16\lambda^{2}_{2}\lambda^{2}_{3}(1-2\lambda^{2}_{0})^{2})}],\\

\langle CHSH \rangle^{2}_{\rho_{AC}}&=&2[(1-2\lambda^{2}_{3})^{2}+4(3\lambda^{2}_{0}\lambda^{2}_{2}+
 \lambda^{2}_{1}\lambda^{2}_{3})\\
 &&+\sqrt{[(1-2\lambda^{2}_{3})^{2}+4(\lambda^{2}_{0}\lambda^{2}_{2}+
 \lambda^{2}_{1}\lambda^{2}_{3})]^{2}-16\lambda^{2}_{0}\lambda^{2}_{2}(1-2\lambda^{2}_{3})^{2})}],\\

\langle CHSH \rangle^{2}_{\rho_{AB}}&=&2[(1-2\lambda^{2}_{2})^{2}+4(3\lambda^{2}_{0}\lambda^{2}_{3}+
 \lambda^{2}_{1}\lambda^{2}_{2})\\
 &&+\sqrt{[(1-2\lambda^{2}_{2})^{2}+4(\lambda^{2}_{0}\lambda^{2}_{3}+
 \lambda^{2}_{1}\lambda^{2}_{2})]^{2}-16\lambda^{2}_{0}\lambda^{2}_{3}(1-2\lambda^{2}_{2})^{2})}].
\end{array}
\end{equation}

First, let us consider the saturation of the inequality (\ref{eq1}).
By optimizing the left hand side of (\ref{eq1}) under the normalization
$\sum^{3}_{i=0}\lambda^{2}_{i}=1$, we have that the upper bound is achieved when $\lambda_{0}\rightarrow -0.423$,
$\lambda_{1} \rightarrow 0.906$, $\lambda_{2} \rightarrow 0$,
$\lambda_{3}\rightarrow 0$, that is
$\langle CHSH\rangle^{2}_{\rho_{AB}}+ \langle CHSH \rangle^{2}_{\rho_{AC}} + \langle CHSH\rangle^{2}_{\rho_{BC}}= 12$.
In this case, there is no violation of the CHSH inequality for any reduced two-qubit density matrices.

Second, let us have a comparison between our trade-off relation and the monogamy relations of CHSH tests \cite{Toner, Marcin, Kurzynski,
P.Kurzynski}. Our trade-off relation gives the restriction on the maximal violations among the
reduced two-qubit systems. An optimal measurement setting which gives rise to the maximal violation for
one pair of reduced density matrix is generally different to that for other pairs of reduced density matrices.
While in the study of monogamy relations, the same measurement settings are applied to the common party of both reduced states.
In \cite{Toner} a monogamy relation has been presented,
$\langle CHSH\rangle^{2}_{\rho_{AC}}+ \langle CHSH \rangle^{2}_{\rho_{BC}} \leq 8$,
which can be violated by some three-quibt states if different measurement settings are allowed to be used
for measuring $\rho_{AC}$ and $\rho_{BC}$ respectively. For instance, for $\lambda_{0}\rightarrow -0.71$ ,$\lambda_{1} \rightarrow 0.69$,
$\lambda_{2} \rightarrow 0.12$, $\lambda_{3}\rightarrow -0.01$, we have $\langle CHSH\rangle^{2}_{\rho_{AC}} =4.15$,
$\langle CHSH\rangle^{2}_{\rho_{BC}} =3.88$, which implies $\langle \langle CHSH\rangle^{2}_{\rho_{AC}}+CHSH\rangle^{2}_{\rho_{BC}} =8.03>8$.

\begin{figure}[htpb]
\renewcommand{\captionlabeldelim}{.}
\renewcommand{\figurename}{Fig.}
\centering\
\includegraphics[width=6.5 cm]{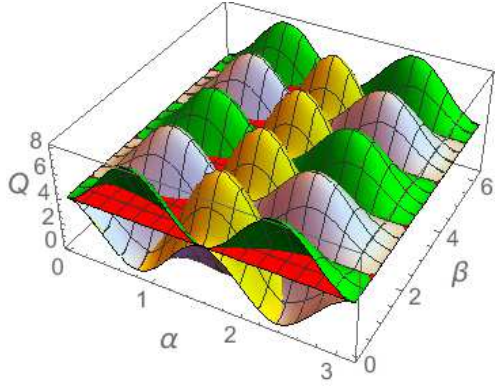}
\caption{{(Color online) \small $Q$ denotes the maximal values of CHSH tests on $\rho_{AB}$ (light-blue
areas), $\rho_{AC}$ (green areas), and $\rho_{BC}$ (yellow areas).
Whenever one of the three pairs of qubits achieves the maximal value
of the CHSH operator, the other two pairs of qubits satisfy the CHSH inequality.}}\label{Fig.1}
\end{figure}

To display the trade-off relations among a three-qubit system, let us set $\lambda_{1}=0$,
$\lambda_{0}=\cos\alpha$, $\lambda_{2}=\sin\alpha\cos\beta$, $\lambda_{3}=\sin\alpha\sin\beta$, where $\alpha \in [0,\pi]$ and
$\beta \in[0,2\pi]$ in (\ref{eq3}).
Fig.\ref{Fig.1} shows the trade off relations among the maximal values of the CHSH tests on pairwise subsystems.
By choosing $\lambda_{0}=\frac{\sqrt{2}}{2}$, $\lambda_{2}=\frac{\sqrt{2}}{2}\cos\theta$ and $\lambda_{3}=\frac{\sqrt{2}}{2}\sin\theta$
we obtain the plane graph Fig. 2. From the Fig. 2 one can see that when one pair of qubits achieves the maximal violation of the CHSH inequality,
the other two pairs of qubits can no longer violate the CHSH inequality.

\begin{figure}[htpb]
\renewcommand{\captionlabeldelim}{.}
\renewcommand{\figurename}{Fig.}
\centering\
\includegraphics[width=6.5 cm]{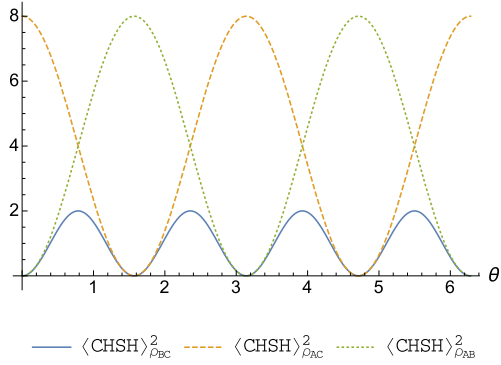}
\caption{{\small  (Color online) The thick line denotes the maximal value of CHSH tests for $\rho_{AB}$,
dashed and dotted lines for $\rho_{AC}$ and $\rho_{BC}$, respectively.}}\label{Fig.2}
\end{figure}

The Theorem can be generalized to arbitrary $n$-qubit systems.
Let $\rho_{A_{1}A_{2}...A_{n}} \in \mathbf{H}^{A_{1}} \otimes
\mathbf{H}^{A_{2}} \otimes \cdots \otimes \mathbf{H}^{A_{n}}$ be an $n$-qubit state.
Let $\rho_{A_{i}A_{j}}\in \mathbf{H}^{ A_{i}}\otimes \mathbf{H}^{ A_{j}}$, $i\neq j$,
be the reduced two-qubit state by tracing over the rest spaces except for the $i$ and $j$-th.

{\bf Corollary}. The maximal values of the CHSH tests on all reduced two-qubit states have following trade-off relation,
\begin{equation}
\sum^{n}_{i<j}\langle CHSH\rangle^{2}_{\rho_{A_{i}A_{j}}} \leq 2n(n-1).
\end{equation}

{\sf Proof}. For any given three-qubit state, say, ${i}\neq{j}\neq{k}$, from Theorem there exists a trade-off relation,
$$
\langle CHSH \rangle^{2}_{\rho_{A_{i}A_{j}}}+\langle CHSH \rangle^{2}_{\rho_{A_{i}A_{k}}}
+\langle CHSH \rangle^{2}_{\rho_{A_{j}A_{k}}} \leq 12=4\binom{3}{2},
$$
where $\binom{m}{n}=m!/(n!(m-n)!)$.
There are $\binom{n}{3}$ tri-qubit subsystems in an $n$-qubit state, which leads to
$$
\sum_{i < j} \langle CHSH \rangle^{2}_{\rho_{A_{i}A_{j}}} \leq 4 \binom{n}{3}\binom{3}{2}=4 \binom{n}{2}=2n(n-1).
$$
\qed

Bell inequalities play important roles in the investigation of quantum nonlocal correlations and quantum entanglement.
By investigating the maximal violations of CHSH inequalities of pairwise sub-qubit systems of a multi-qubit
system, we have presented a trade-off relation among these pairwise violations.
The trade-off relation gives rise to restrictions on the distribution of non-locality among the
subsystems. It implies that for a three-qubit system, it is impossible that all pair of qubits states
violate the CHSH inequality simultaneously. And if one of the three pairs of qubits violates the CHSH inequality
maximally, the other two pairs of qubits must both obey the CHSH inequality.
Moreover, this trade-off relation could be also used to quantify some kinds of genuine three-qubit quantum non-locality
when each pair of qubit states admits LHV models.
Here it should be noted that, since the reduced two-qubit states are mixed ones, the CHSH inequality is neither
necessary nor sufficient to verify the nonlocality. Other Bell inequalities based trade-off relations are also
desired for investigation of non-locality distributions.

{\it Acknowledgements.} This work is supported by the NSFC under number 11275131.
Qin acknowledges the fellowship support from the China scholarship council.

\end{document}